\documentclass[10pt,letterpaper,onecolumn]{article}

\usepackage{graphicx}
\usepackage{times, amssymb, amsmath, amsthm, latexsym, subfigure, mathrsfs, multirow,url}
\usepackage[letterpaper,footskip=0.5in,paperwidth=8.5in,paperheight=11in,centering,vmargin={1in,1in},hmargin={1in,1in}]{geometry}
\usepackage[round,sort&compress]{natbib}
\usepackage[vlined,algoruled,titlenotnumbered]{algorithm2e}

\usepackage{pictex}
\def \color#1]#2{}

\newtheorem{theorem}{Theorem}

\newcommand{\hide}[1]{}

\renewcommand{\citeyear}[1]{\citeyearpar{#1}}
\newcommand{\nodes}{\mathcal{N}}
\newcommand{\edges}{\mathcal{E}}
\newcommand{\hgraph}{(\nodes, \edges)}
\newcommand{\forest}{\mathcal{T}}

\newcommand{\QED}{\hfill $\Box$}
\newcommand{\routine}[1]{\texttt{\small #1}}
\newcommand{\treeConvex}{\routine{isTreeConvex}}
\newcommand{\isAcyclic}{\routine{isAcyclic}}
\newcommand{\genForest}{\routine{genForest}}
\newcommand{\treeTest}{\routine{treeTest}}

\newcommand{\defKeywords}{
\SetKw{Or}{or} \SetKw{Ret}{return} \SetKw{Exit}{exit}
\SetKw{False}{false} \SetKw{True}{true} \SetKw{Not}{not}
\SetKw{AND}{and}
\SetKw{Or}{or} \SetKw{Break}{break}
\SetKwData{Consistent}{consistent} \SetKwData{Stack}{s}
\SetKwData{Head}{head} \SetKwData{Tail}{tail} \SetKwData{Max}{max}
\SetKwData{Min}{min}
\SetKwFunction{Pred}{pred} \SetKwFunction{Succ}{succ}
\SetKwFunction{Eliminate}{eliminate} \SetKwFunction{Pop}{pop}
\SetKwFunction{Push}{push} \SetKwFunction{Empty}{empty}
\SetKwFunction{Solve}{solve} \SetKwFunction{SlowComp}{compose}
\SetKwFunction{FastComp}{fastCompose}
\SetKwFunction{Disjoint}{disjoint}
\SetKwFunction{RemoveValues}{removeValues} \SetArgSty{} }

\newcommand{\fun}[2]{
\hskip 1em
\Indm \Indm
\FuncSty{#1} \;
\Indp \Indp

#2
}

\begin{document}
\title{A Survey of Tree Convex Sets Test}
\author{Yuanlin Zhang and Forrest Sheng Bao \\ Dept. of Computer Science, Texas Tech University \\Lubbock, Texas 79409
\thanks{yzhang@cs.ttu.edu, forrest.bao@gmail.com}}
\date{}
\maketitle

\begin{abstract}
Tree convex sets refer to a collection of sets such that each set in
the collection is a subtree of a tree whose nodes are the elements of
these sets. They extend the concept of row convex sets each of which
is an interval over a total ordering of the elements of those sets.
They have been applied to identify tractable Constraint Satisfaction
Problems and Combinatorial Auction Problems. Recently, polynomial
algorithms have been proposed to recognize tree convex sets. In this
paper, we review the materials that are the key to a linear
recognition algorithm.
\end{abstract}

\section{Introduction}
Given a set $U$, a collection $S$ of subsets of $U$ is tree convex
%Tree convex sets are introduced in \citep{ZY03b}.
if there exists a tree $T$ with nodes $U$ such that every set of $S$
is a subtree \citep{ZY03b} of $T$. Row convex sets are a collection
of sets that are tree convex with respect to a chain (a special tree
with nodes $U$). Row convex sets correspond to another well studied
concept: {\em consecutive ones property} of matrices. Let $M$ be the
matrix whose rows are indexed by the elements of $S$ and columns
indexed by those of $U$ in terms of a total ordering over $U$. An
entry of $M$, indexed by $(s,a)$ with $s \in S$ and $a \in U$, is
one if and only if $a \in s$. $M$ has consecutive ones property
\citep{FulkersonG65} with respect to its rows if there is a total
ordering of $U$ such that the ones on each row is consecutive.
Clearly, the sets of $S$ are row convex if and only if the matrix
$M$ has consecutive ones property.

The property of tree convex and row convex sets has been employed to
identify tractable Constraint Satisfaction Problems (CSP). CSP
problems have found many successful applications in Artificial
Intelligence and Combinatorial Problems \citep{Dec03}. However, in
general, CSP problems are NP-hard. Continuous research effort has
been made to identify tractable CSP problems.
% i.e., problems that can be solved in polynomial time.
An important approach is to make
use of semantic properties of the constraints. For {\em monotone
constraints}, path consistency implies global consistency
\citep{Mon74}.  \citet{vBD95} generalize
monotone constraints to a larger class of {\em row convex
constraints} which is in turn expanded to {\em tree convex
constraints} by \citet{ZY03b}. The tractability of
these constraints results from the nice intersection property of
tree convex constraints.

Recently, tree convex sets also have found applications in
combinatorial auctions. Given a set $U$ of items and a collection of
bids each of which is a subset of $U$, the problem to decide the
winners is NP-complete \citep{RothkopfPH98} in general. However, when
the collection of bids are tree convex, the problem becomes
tractable \citep{SandholmS03}. (Note that although ``tree convexity"
is not used in that paper, the concept there is exactly the same as
tree convexity.)

An interesting and challenging question raised in the application of
tree convex sets in both CSP and Combinatorial Auctions is how
efficiently one can test the tree convexity of a given collection of
sets. There is abundant related research work under the umbrella of
{\em consecutive ones property test}, i.e., row convexity test. The
consecutive ones problem was first proposed by 
\citet{FulkersonG65}. A linear algorithm was then developed by
\citet{BoothL76}. It uses quite complex data
structures and involved techniques. There exists continuous work,
e.g., by \citet{MeidanisPT98}, \citet{HabibMPV00}, and \citet{Hsu02}, to improve the
understanding of consecutive ones property and its test. For tree
convexity test, polynomial algorithms have been recently designed by
\citet{Yos03} and \citet{ConitzerDS04}. Yosiphon makes use of complex data
structures and ideas inherited from consecutive ones property work.
The resulting algorithm is rather involved and has a complexity of
$O(mn)$. Conitzer et al. proposes a ``simple" algorithm but with a
still very high time complexity $O(mn^2)$ where $m$ is the number of
sets (bids) and $n$ the number of all distinct elements in the sets,
i.e., the number of all items to bid.

A very interesting question is whether there are linear algorithms
for tree convexity test like row convexity test. In fact, it is
listed as one of the open questions in \citep{ConitzerDS04}. This
question can be answered positively if we take the collection of
sets as a hypergraph. With this perspective, we are not only able to
identify a simple and nice characterization of tree convex sets
using hypergraphs and properties of hypergraphs, but also to connect
this problem with the long line research of conjunctive query
evaluation in databases and tree decomposition in Constraint
Satisfaction Problems \citep{BeeriFMY83,DP89,GottlobS08}. As a
result, an existing simple and elegant linear algorithm for
hypergraphs by \citet{TarjanY84} can be directly
used to test tree convexity.

Due to a well known example in Constraint Satisfaction Problems
where an optimal algorithm AC-4 on enforcing arc consistency does
not perform better than a non-optimal algorithm AC-3 \citep{Wal93} in
most cases, we also carry out experiments on a set of randomly
generated problems to compare the linear algorithm with the one in
\citep{ConitzerDS04}. Experimental results show that the former is
significantly faster than the latter.

Section~\ref{sec:background} reviews basic concepts and terms
including those that might have different meanings in different
context. The details of a characterization of tree convex sets and
related work are given in Section~\ref{sec:characterization}. To
make this survey self contained, a test algorithm including Tarjan
et al.'s algorithm is presented in Section~\ref{sec:algorithm}.
Experimental results are given in Section~\ref{sec:experiments}
before we conclude the paper.

%to Helly
% constraints for which path consistency ensures global consistency.

\section{Background}
\label{sec:background}

% basics of graph
In this section, we will review the basics of tree convex sets, the
related concepts of graphs and hypergraphs, and some applications of
tree convex sets in Constraint Satisfaction Problems and
Combinatorial Auction problems.

A {\em graph} is a tuple $(N,E)$ where $N$ and $E$ are sets,
elements of $N$ are called vertices or nodes and those of $E$ edges,
and each edge is a set of at most two vertices. Hypergraphs
generalize graphs by allowing an edge to be a set of arbitrary
number of vertices. Specifically, a {\em hypergraph} $H$ is a pair
$\hgraph$ where $\nodes$ is a set of vertices, and $\edges$ consists
of nonempty subsets of $\nodes$ that are called {\em hyperedges}.
Berge's book \citeyearpar{Berge73} is an excellent reference for
hypergraphs.
% Note that both $E$ of a graph and $\edges$ of a
% hypergraph should be taken as {\em multisets}.

\subsection{Notations and results in graphs}

A {\em clique} of a graph is a set of pairwise adjacent vertices. A
graph is {\em chordal} if every cycle of length at least four has a
chord, i.e., an edge joining two nonconsecutive vertices on the
cycle. {\em Forests, trees, chains} and {\em (simple) path} are
defined as usual. To reduce the potential confusion or
misunderstanding, we repeat the following definitions. A graph
$(N_1, E_1)$ is a {\em subgraph} of $(N, E)$ if $N_1 \subseteq N$
and $E_1 \subseteq E$. Given a tree, a {\em subtree} is defined as a
connected subgraph of the tree. A {\em forest on a set $S$} is a
forest whose vertex set is exactly $S$.
%We also call a set
%$I$ a {\em subtree} of a forest \forest\ if there exists a subtree
%of some tree in \forest\ such that its vertex set is exactly $I$. An
%empty set is a {\em subtree} of any forest.

\subsection{Notations and results in hypergraphs}

We introduce in this section dual hypergraphs, acyclic hypergraphs,
join trees and some results on hypergraphs. Throughout this paper,
we may use ``graphs" for ``hypergraphs" and ``edges" for
``hyperedges" when their meaning is clear from the context.

The {\em graph $G(H)$} of a hypergraph $H$ is the graph whose
vertices are those of $H$ and whose edges are pairs $\{x,y\}$ such
that $x$ and $y$ are in a common edge of $H$. A hypergraph $H$ is
{\em conformal} if every clique of $G(H)$ is contained in an edge of
$H$.

The {\em dual graph $H^*$} of a graph $H=(\{v_1, v_2, \ldots, v_n\},
\{S_1, S_2, \ldots, S_m\})$ is a hypergraph $(\{S_1, S_2, \ldots,
S_m\}$, $\{R_1, R_2, \ldots, R_n\})$ where for $i \in 1..n$, $R_i =
\{S_j ~|~ v_i \in S_j, j \in 1..m\}$. The edge $R_i$ is the set of
edges of $H$ that involve vertex $v_i$. Intuitively, one can take
$R_i$ as $v_i$.

     The acyclicity of a hypergraph involves a sequence of concepts
     defined below.
     $H$ is {\em reduced} if no edges of it properly contain another
     edge and every  node is in some edge. The {\em reduction} of $H$ is $H$
     with any contained edges and non-edge nodes removed.

     Let $H = \hgraph$ be a hypergraph with nodes $x$ and $y$ in $\nodes$.
     A {\em path} from $x$ to $y$ in $H$ is a sequence of edges
     $E_1, E_2, \ldots, E_k$ ($k \ge 1$), such that $x \in E_1$, $y \in E_k$
     and $E_i \cap E_{i+1} \neq \emptyset$ for $i \in [1..k-1]$.
     $E_1, E_2, \ldots, E_k$ is also called a path from $E_1$ to $E_k$.

     Two nodes (or edges) are {\em connected} if there is a path
     between them. A set of edges is {\em connected} if every pair
     of the edges is connected. A {\em connected component} of $H$ is a
     maximal connected set of edges.

     Given a hypergraph and a subset of its nodes, we will now
     define the ``projection" of the graph on these nodes. Let $M$ be
     a set of nodes of the hypergraph $\hgraph$. The {\em set of
     partial edges generated by $M$} is defined to be the reduction
     of $\{E \cap M ~|~ E \in \edges \} - \{\emptyset\}$. It is also called
     a {\em node-generated set of partial edges}. Given a
     set of edges $\mathcal{F}$, we say $(E,F)$, where $E,F \in
     \mathcal{F}$, is an {\em articulation pair} if $E\cap F$ is an
     {\em articulation set}, i.e., removing $E \cap F$ from every
     edge in $\mathcal{F}$ strictly increases the number of
     connected components of $\mathcal{F}$.

     A {\em block} of a reduced hypergraph is a connected
     node-generated set of partial edges without articulation set.
     A reduced hypergraph is {\em acyclic} if all its blocks
     have less than two edges. A hypergraph is said to be {\em acyclic}
     if its reduction is.

\begin{figure}[htb!]
        \begin{center}
        \input{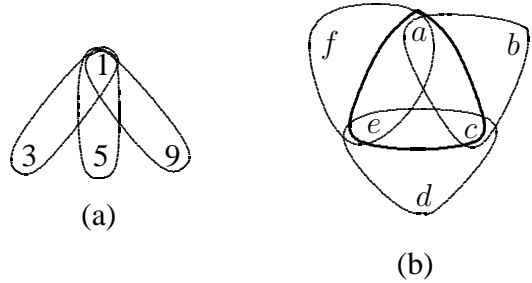}
   \end{center}
   \caption{\label{fig:fig1} Acyclic graphs can be either tree convex or non tree convex. The letters are the vertices
   and the edges are represented by enclosed curves.}
   \end{figure}

    As examples, consider the graphs in Figure~\ref{fig:fig1}(a) and Figure~\ref{fig:fig1}(b).
    The former is acyclic, following our intuition. However, the latter is also acyclic.
    Although $a,e,c,a$ form a ``cycle," the graph is acyclic by definition because they the cycle
    is covered by the edge $\{a, e, c\}$.

     We define join tree below.
     Given a collection $S$ of sets: $S=\{S_1, S_2, \ldots, S_m\}$,
     the {\em intersection graph} for $S$,
     denoted $I_S$, is the undirected graph $(S, E)$ where $\{S_i, S_j\} \in
     E$ iff $S_i \cap S_j \neq \emptyset$. A path $S_{i_1}, S_{i_2},
     \ldots, S_{i_k}$ of $I_S$ is an {\em $A$-path} if $A \in S_{i_j} \cap
     S_{i_{j+1}}$ for all $j \in 1..k-1$. A subgraph $G=(S, E')$ of
     $I_S$ is a {\em join graph} if for every pair of nodes $S_i$
     and $S_j$ of $S$ and every $A \in S_i \cap S_j$, there is an $A$-path
     from $S_i$ to $S_j$ in $G$. A {\em join tree} is a join graph that is
     a tree. A hypergraph $\hgraph$ has a join tree if there is a join tree for $\edges$.
     Acyclic graphs and join trees are closely related as revealed by the following
     result.
     \begin{theorem}[\citep{BeeriFMY83}] \label{th:acyclicGraph}
     The following statements on hypergraph $H$ are equivalent:
     \begin{itemize}
        \item $H$ is acyclic.
        \item $H$ has a join tree.
        \item $H$ is conformal, and $G(H)$ is chordal.
     \end{itemize}
     \end{theorem}

\subsection{Tree convex sets}
A collection of sets $S_1, S_2, \cdots, S_m$ is {\em tree convex
with respect to a forest $\forest$ on $\cup_{i \in 1..m}
S_i$} if every $S_i$ is a subtree of $\forest$. For example, the
sets $\{a,b,c\}$, $\{a,b,d\}$, and $\{a,c,d\}$ are tree convex with
respect to the tree with vertices $\{a,b,c,d\}$ and edges
$\{\{a,b\}, \{a,c\}, \{a,d\}\}$.

% Row convex sets ?? and consecutive ones property ?? references on
% C1P problem ??

\subsection{Tree convex constraints and problems}
A binary {\em constraint network} consists of a set of variables $V
= \{x_1, x_2, \cdots, x_n\}$ with a finite domain $D_i$ for each
variable $x_i \in V$, and a set of binary constraints $C$ over the
variables of $V$. $c_{xy}$ denotes a constraint on variables $x$ and
$y$ which is defined as a relation over $D_x$ and $D_y$. Operations
on relations, e.g., {\em intersection ($\cap$), composition
($\circ$), and inverse}, are applicable to constraints. The {\em
arc} and {\em path} consistency are defined as in \citep{Mac77}, and
global ($k$ consistency) consistency in \citep{Fre78}.

Given a constraint $c_{xy}$, the {\em image} of a value $a$ of $x$
is the set of values of $y$ that are compatible with $a$ under
$c_{xy}$. A constraint $c_{xy}$ is {\em tree convex with respect to
a forest $\forest$ on $D_y$} if the images of all values of $D_x$
are tree convex with respect to $\forest$. A constraint network is
{\em tree convex} if there exists a forest on the domain of each
variable such that every constraint $c_{xy}$ of the network is tree
convex with respect to the forest on $D_y$.

If a tree convex constraint network is arc and path consistent, it
is global consistent \citep{ZY03b}, which implies that a solution can
be found in polynomial time.

\subsection{Combinatorial auction problems}

Emerging as key mechanisms for allocating goods, tasks, resources
etc., combinatorial auctions \citep{CramtonSS06} allow the bidders to
bid on bundles of items, instead of single item. The problem to
determine the winners in combinatorial auctions is NP-complete
\citep{RothkopfPH98}. However, restricted classes of combinatorial
auction problems have been identified. For those classes, there
exist efficient polynomial algorithms. We are particularly
interested in the class of problems where an item graph of the bids
is a tree \citep{ConitzerDS04}.

Every {\em bid} is a set of items. Given a combinatorial auction
clearing problem instance (i.e., a set of bids), the graph $G =
(I,E)$, where $I$ corresponds to the items in the instance, is a
\emph{(valid) item graph} if for every bid, the set of items in that
bid constitutes a connected subgraph of $G$. $G$ is a {\em item
tree} if it is a tree.

It is straightforward to verify, by the definitions, that a set of
bids is tree convex iff there is an item tree for the bids.

Conitzer et al. % \citeyear{ConitzerDS04})
proposed an algorithm to recognize tree convexity with complexity of
$O(mn^2)$ where $m$ is the total number of bids and $n$ the number
of total items in the auction. Given a collection of bids $S=\{S_1,
S_2, \dots, S_m\}$, the algorithm first constructs a graph with
vertices $\cup S ( = S_1 \cup S_2 \cup \cdots \cup S_m)$, and
weighted edges $G=\{(\{a,b\}, weight) ~|~ \exists s \in S \mbox{
such that } a,b \in s, \mbox{ and } weight =|\{s \in S~:~ a,b \in
s\}|\}$. It next finds the maximum spanning tree $T$ of $G$.

The sets of $S$ are tree convex iff the sets are tree convex with
respect to $T$ \citep{ConitzerDS04}.

\section{Characterization of tree convex sets}
\label{sec:characterization}

   Given a collection of sets $S= \{S_1, S_2, \dots, S_m\}$, let
   $U(S) = \cup_{s \in S}s$. The {\em hypergraph of $S$} is $(U(S), S)$.
   The {\em dual hypergraph} of $S$ is the dual graph of $(U(S), S)$.

   To identify whether $S$ is tree convex, one convenient way is to
   look at the hypergraph of $S$. Consider the example $\{ \{1, 3\}, \{1, 5\},
   \{1,9\}\}$ in Figure~\ref{fig:fig1}(a). Clearly, its hypergraph is acyclic and suggests
   a tree with respect to which the collection is tree convex.
   However, we have the following observations
   about the relationship between a collection of sets and the acyclicity
   of their hypergraphs.

\begin{figure}[htb!]
        \begin{center}
        \input{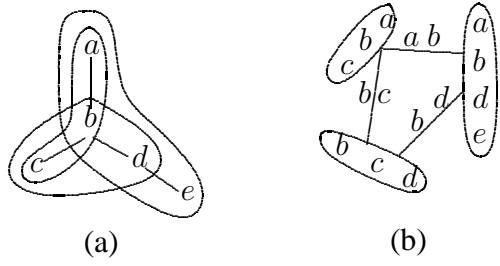}
   \end{center}
   \caption{\label{fig:fig2} Tree convex sets might not be acyclic. (a) Straight lines represent
   edges of the underlying tree on the vertices. (b) Enclosed curves represent nodes which correspond to
   edges in (a). Letters on the straight edges represent the intersection of the nodes at their ends.}
   \end{figure}

   The graph of $S$ is acyclic does not necessarily mean the tree convexity
   of $S$. In other words, the graph of a non tree convex sets could be acyclic.
   Consider the collection $S= \{ \{a, e, f\}, \{c, d, e\}, \{a, b, c\}, \{a, c, e\}\}$ in
   Figure~\ref{fig:fig1}(b). As mentioned before, $S$ is acyclic. However, it is not tree convex.
   Assume otherwise it is tree convex with
   respect to a tree $T$. There are paths on $T$: $P1: a \rightarrow c$ (because $a,b$, and $c$ form a subtree of $T$),
   $P2: c \rightarrow e$,
   $P3: e \rightarrow a$. Clearly,
   $P1P2P3$ forms a cycle, a contradiction to the fact that $T$ has no cycles.
   Another observation is that not all tree convex sets form an acyclic hypergraph.
   The example $S = \{\{a,b,c\}, \{a, b, d, e\}, \{b, c, d\}\}$ (Figure~\ref{fig:fig2}(a)), given by
     Yosiphone\footnote{Personal communication 2004.}, is tree convex but not acyclic. Each set
     of $S$ is a subtree of the tree shown in the figure.
     From the intersection
     graph of $S$ in Figure~\ref{fig:fig2}(b), there does not
     exist a join tree for $S$. So, $S$ is not acyclic.

In fact, the tree convexity of a collection is related to the
acyclicity of its dual graph.

\begin{theorem} \label{th:treeConvex}
     A collection $S$ of sets is tree convex iff its dual hypergraph
     is acyclic.
\end{theorem}

\proof{

    Given a collection $S$ of sets, let $H=(\{v_1, v_2, \ldots, v_n\}, S)$ be its hypergraph.
    Here we take $U(S)$ as $\{v_1, v_2, \ldots, v_n\}$.
    Let $D = (S, \{R_1, R_2, \ldots, R_n\})$ be the dual graph of S.
%    By definition of dual graph, there is a one-one correspondence
%    between $U(S)$ and the edges of $D$.
%    Every edge of $D$ can be taken as a vertex of $H$.

    Necessary condition. Let $\mathcal{T}$ be a tree on $U(S)$ such that $S$ is tree
    convex  with respect to it. The idea is to construct a join tree
    for $D$ so that $D$ is acyclic by Theorem~\ref{th:acyclicGraph}.
    We now construct a tree $\mathcal{T}'$=$(V, E)$ where $V=\{R_1, R_2, \ldots, R_n\}$.
    For all $R_i, R_j \in V$, $\{R_i, R_j\} \in E$ if and only if $\{v_i, v_j\}$ is an edge of
    $\mathcal{T}$. We next show that $\mathcal{T}'$ is a join tree for $D$. Consider
    any two vertices $R_i$ and $R_j$ such that $R_i \cap R_j \neq \emptyset$ and any
    $f \in R_i \cap R_j$ (note $f$ is an edge of $H$).
    By definition of dual graph, $v_i, v_j \in f$ because $f \in R_i \cap
    R_j$ and $R_i$ and $R_j$ consist of edges involving $v_i$ and $v_j$ respectively. There is a unique path
    from $v_i$ to $v_j$ in $\mathcal{T}$. Let it be $P=v_i, v_{i+1}, \ldots, v_{j}$.
    $S$ is tree convex implies $f$ is a subtree of $\mathcal{T}$. Since both $v_i$ and $v_j$
    belong to $f$, all vertices on
    $P$ are in $f$. Corresponding to $P$, there is a path $P'=R_i, R_{i+1}, \ldots, R_{j}$
    in $\mathcal{T}'$ by the construction of $\mathcal{T}'$.
    For all $k \in i..j$, since $v_k \in f$, we have $f \in R_{k}$. Hence, $P'$ is an $f$-path from
    $R_i$ to $R_j$. Therefore, $\mathcal{T}'$ is a join tree of $D$.

    Sufficient condition. Since the dual graph of $S$ is acyclic,
    there is a join tree $\mathcal{T}'=(\{R_1, R_2, \ldots, R_n\}, R)$ for $D$ by Theorem~\ref{th:acyclicGraph}.
    We will show that there is a tree $\mathcal{T}$ under which $S$ is tree convex.
    Construct $\mathcal{T} = (\{v_1, v_2, \ldots, v_n\}, E)$ where $(v_i,v_j) \in E$ if and only if
    $\{R_i, R_j\} \in R$. Clearly, $\mathcal{T}$ is a tree. We next prove that for any $s \in S$,
    $s$ is a subtree of $\mathcal{T}$.
    Specifically, we show that for any two vertices $v_i$ and $v_j$ of the edge $s$, there exists
    a path from $v_i$ to $v_j$ in $\mathcal{T}$ and the nodes on the path are in $s$.
    By definition of dual graphs, $s \in R_i$ and
    $s \in R_j$ because $v_i, v_j \in s$. Since $\mathcal{T}'$ is a join tree of $D$, there is an $s$-path from $R_i$
    to $R_j$: $R_i, R_{i+1}, \ldots, R_j$ in $\mathcal{T}'$. By the
    construction of $\mathcal{T}$, $v_i, v_{i+1}, \ldots, v_j$ is a path of
    $\mathcal{T}$. For all $k \in 1..j$, since $s \in R_k$, we have $v_k \in s$.
    Hence, $s$ is a subtree of $\mathcal{T}$ and thus $S$ is tree
    convex. \QED
}

\begin{figure}[htb!]
        \begin{center}
        \input{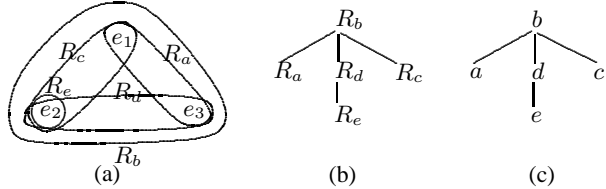}
   \end{center}
   \caption{\label{fig:fig3} (a) The dual graph of $S$. Every edge has a label of $R$ with subscript.  (b) A join tree.
   (c) Tree derived from (b). The nodes are the elements in the original sets.}
   \end{figure}

To illustrate the concepts used in the proof, consider the
collection $S$ = $\{\{a,b,c\}$, $\{a, b, d, e\}, \{b, c, d\}\}$
again. Let $e_1 = \{a,b,c\}, e_2 = \{a, b, d, e\}$, and $ e_3 = \{b,
c, d\}$. The hypergraph of $S$ is $H = (\{ a, b, c, d, e\}, \{e_1,
e_2, e_3\})$ (Figure~\ref{fig:fig2}(a)). The dual graph of $S$ is $D
= (\{e_1, e_2, e_3\}, \{R_a, R_b, R_c, R_d, R_e \})$
(Figure~\ref{fig:fig3}(a)) where $R_a=\{e_1, e_2\}, R_b=\{e_1, e_2,
e_3\}, R_c=\{e_1, e_3\}, R_d=\{e_2, e_3\}, R_e=\{e_2\}$. Since $R_e$
is a subset of $R_d$ and other edges are subsets of $R_b$, we have a
join tree shown in Figure~\ref{fig:fig3}(b). So, $D$ is acyclic.
From the join tree, we can construct a tree on the nodes of the
original sets as in Figure~\ref{fig:fig3}(c). $S$ is tree convex
with respect to the tree.

A result similar to Theorem~\ref{th:treeConvex} was discovered by
 \citet{GoodmanS83} long time ago in the study of
database schemas. They provided a rather comprehensive
characterization of acyclic hypergraphs. One of their main results
is the relationship between acyclic hypergraph and chordality and
conformality which is well known by the constraint community
\citep{BeeriFMY83,Dec03}. However, another result is not known well
but directly related to the characterization of tree convexity.  It
is worth reviewing the result here. First, we introduce some of
their terms that are not well known in the constraint community. In
the case that confusion could arise from the use of common
terminologies, we underline the terms.

Given a hypergraph $H = \hgraph$, a {\em \underline{dual graph}} for
$H$ \citep{GoodmanS83} is a graph $G = (V_\edges, F)$ equipped with a
one one onto map $V_\edges$ to $\edges$ indicating which node of $G$
represents which edge of $\edges$. Note that $G$ is not a hypergraph
here, but just a graph. One type of \underline{dual graph} used by
Goodman and Shmueli is an {\em \underline{intersection graph}},
denoted by $\Omega(H)$. $\Omega(H) = (V_\edges, F)$ such that
$\{x,y\} \in F$ iff $E_x \cap E_y \neq \emptyset$ where $E_x$ and
$E_y$ are the edges (of $H$) represented by $x$ and $y$
respectively. A second type of \underline{dual graph} is a qual
graph \citep{BernsteinG81}. Given $u \in \nodes$, the {\em dual} of
$u$ is $u^* = \{E \in \edges ~|~ u \in E\}$. A {\em qual graph} for
$H$ is any \underline{dual graph} $G = (V_\edges, F)$ such that for
each $u \in \nodes$, the subgraph of $G$ induced by nodes
representing elements of $u^*$ is connected. One can verify that the
graph of Figure~\ref{fig:fig3}(c) is a qual graph of the hypergraph
of Figure~\ref{fig:fig3} (a). The nodes $a$ to $e$ of
Figure~\ref{fig:fig3}(c) represent edges $R_a$ to $R_e$. As an
example, consider node $e_3$. Its dual $e_3^* = \{R_c, R_b, R_d\}$.
The subgraph of Figure~\ref{fig:fig3}(c) induced by $a,b,c$
(representing the elements of $e_3^*$) is connected.

A database {\em schema} can be thought of as a hypergraph whose
nodes are the schema's attributes and whose edges are the schema's
relations. A hypergraph $H$ is a {\em tree schema} if some qual
graph for it is a tree.

Now we are ready to present Goodman and Shmueli's result
\cite[Theorem 6]{GoodmanS83}.

\begin{theorem}[\citealt{GoodmanS83}]
\label{th:qualgraph}
A hypergraph $H$ is a tree schema iff $H$ is acyclic.
\end{theorem}

Theorem~\ref{th:treeConvex} and \ref{th:qualgraph} are equivalent.
First, One can show that if a collection of sets is tree convex with
respect to a forest, it is tree convex with respect to a tree, and
vice versa. Next, by Theorem~\ref{th:treeConvex}, hypergraph $H$ is
acyclic iff the collection of the edges of its dual graph, $H^*$, is
tree convex. Thirdly, a key observation is that the collection of
edges of $H^*$ is tree convex iff some qual graph for $H$ is a tree.
By the definition of tree convexity, the former condition holds iff
there exists a tree $T$ with nodes of $H^*$ such that every edge of
$H^*$ is a subtree of $T$. Clearly, by the definition of qual graph,
$T$ is a qual graph for $H$. Finally, by definition of tree schema,
$H$ is a tree schema iff there exists a qual graph for $H$.

Recently, a nice and more general result on hypergraphs was
discovered by Gottlob and Greco \citep{GottlobG07}.

\begin{theorem}[\citealt{GottlobG07}] \label{th:hyperTree}
Let $k$ be a number and $H = \hgraph$ a hypergraph such that for
each node $v \in \nodes$, $\{v\} \in \edges$. Then, a $k$-width tree
decomposition of an item graph for $H$ exists if and only if $H^*$
has a $(k+1)$-width strict hypertree decomposition.
\end{theorem}

Essentially, the hypergraph $H$ is a set of bids (i.e., a collection
of sets). A detailed explanation of the concepts of {\em $k$-width
tree decomposition of a graph} and {\em $(k+1)$-width (strict)
hypertree decomposition of a hypergraph} can be found in
\citep{GottlobG07}. This result relates a more general property of a
hypergraph with some property of the its dual. A $1$-width tree
decomposition of an item graph for $H$ exists if and only if an item
graph for $H$ is a tree, i.e., $H$ is tree convex. By definition of
strict hypertree decomposition, one can show that a hypergraph has a
2-width strict hypertree decomposition if and only if it is acyclic.
So, Theorem~\ref{th:hyperTree} implies Theorem~\ref{th:treeConvex}
and thus \ref{th:qualgraph}.

% ??? -- revise the rest to make sure no new algorithm or our
% algorithm appear

{\bf Remark.} Given a hypergraph $H$ (representing the topological
structure of a CSP problem), its \underline{\underline{dual
(constraint) graph}} is defined as the \underline{intersection
graph} for $H$ in \citep{Dec03}. Clearly, the dual graph is different
from \underline{dual graph} and \underline{\underline{dual
(constraint) graph}}. The definition of intersection graph agrees
with that of \underline{intersection graph}. As for the definitions
of acyclic graphs, we follow those in \citep{BeeriFMY83}. Acyclic
hypergraphs are called hypertrees in \citep{Dec03}, but
$\alpha-$acyclic graphs in \citep{Fagin83} where other types of
acyclicity are also introduced.
% Also, $G(H)$ is called {\em primal graph} of $H$ in \citep{Dec03}.

\section{Algorithms to identify tree convexity}
\label{sec:algorithm}
   By Theorem~\ref{th:treeConvex}, we have the following algorithm to test
   the tree convexity of a given collection $S$ and produce a tree if the given collection is
   tree convex.

\begin{algorithm}[htb!]
\setlength{\algomargin}{3em}

\defKeywords

\dontprintsemicolon
   \fun{\treeConvex({\bf in} $S$)}{
    \nl Let $D$ be the dual graph of $S$ \;
    \nl \If{\isAcyclic($D$, $R$, $\gamma$)} {
    \nl     \genForest($D$, $R$, $\gamma$, $\mathcal{T}$) \;
    \nl     \Ret (\True, $\mathcal{T}$) \;
        } \Else {
    \nl     \Ret \False
        }
   }
\caption{\Indp \scriptsize Recognize tree convexity of sets}
\label{alg:treeConvex}
\end{algorithm}
The algorithm first constructs the dual graph $D$ of $S$. The
function \isAcyclic($D$, $R$, $\gamma$) returns true and data
structures $R$ and $\gamma$ (discussed below) if the graph of $D$ is
acyclic, and it returns false otherwise. In the former case, using
$R$ and $\gamma$, \genForest($D$, $R$, $\gamma$, $\mathcal{T}$)
builds tree $\mathcal{T}$ (using $R$ and $\gamma$) with respect to
which $S$ is tree convex.

Based on the work by \citet{RoseTL76}, \citet{TarjanY84} proposed a simple linear algorithm
(maximum cardinality search) to identify whether a hypergraph is
acyclic. Although maximum cardinality search on a graph can be
easily found in a wide range of references \citep{Dec03}, very few
references involve the search over hypergraphs. We include it here
to make our presentation complete, with the correction of some
errors in the original presentation.

Given a graph $\hgraph$, the key behind this algorithm is to compute
three mappings $\alpha$, $\beta$,and $\gamma$. A mapping is a
(possibly partial) function that assigns a node and/or an edge to a
number between (including) $1$ and $|\nodes|$. Specifically, the
domain of $\alpha$ is $\nodes$, that of $\beta$ is $\nodes$ and
$\edges$, and that of $\gamma$ is $\edges$. The algorithm, called
{\em restricted maximum cardinality search on hypergraph}, works as
follows. It first selects an edge $s$ from $\edges$ arbitrarily.
Mapping $\alpha$ assigns the nodes of $s$ the number from $n$ to
$n-|s|+1$ one by one. An edge is {\em exhausted} if all of its nodes
have been assigned a number by $\alpha$, and {\em nonexhausted}
otherwise. Next we select a nonexhausted edge $t$ with the maximum
number of nodes assigned by $\alpha$ (tie will be broken
arbitrarily). Let $n_1$ be the largest number that is smaller than
$|\nodes|$ but not used by $\alpha$ yet. Assign the non-assigned
nodes of $t$ to numbers from $n_1$ to $n_1 - |t|+1$. Repeat this
process until every node of the graph is assigned a number by
$\alpha$. $R(i)$ is used to remember the $i^{th}$ selected edge. The
mapping $\beta$ is defined as follows. If $s$ is the $i^{th}$
selected edge, $\beta(s)=i$. Otherwise, it is not defined. For a
node $v$, $\beta(v)$ is defined as $\beta(s)$ where $s$ is the first
selected edge such that $v \in s$, i.e., $\beta(v) = \min\{\beta(s)
~|~ s \mbox{ is selected and } v \in s\}$. (Note that in line 12 of
the algorithm, $\beta(E) \leftarrow k$ is redundant. We keep it
there to make it compatible with the original algorithm. It also
makes the definition of $\beta$ clearer.) For each edge $s$, if $s$
is not selected during the process, $\gamma(s)$ is $\beta(v)$ where
$v \in s$ is the last one to be assigned a number by $\alpha$, i.e.,
$\gamma(s) = \max\{\beta(v)~|~ v \in s\}$; if $s$ is selected by the
process, $\gamma(s)$ is $\beta(v)$ if $v \in s$ is the last node
assigned by $\alpha$ strictly before $s$ is selected, i.e.,
$\gamma(s) = \max\{\beta(v)~|~ v \in s \mbox{ and } \beta(v) <
\beta(s)\}$, in the last case, if $\beta(v) = \beta(s)$ for all $v
\in s$, $\gamma(s)$ is not defined.

The mappings are then employed to test the acyclicity of a graph. Given a
hypergraph $H$, assume totally $k$ edges are selected during the
process above. $H$ is acyclic iff for each $i \in 1..k$ and each
edge $s$ such that $\gamma(s)=i$, $s \cap \{v ~|~ \beta(v) < i\}
\subseteq R(i)$. The code from line 26 to 32 implements this test.
% During the test of $R(i)$, the data structure $index(v) = i$ if $v
% \in R(i)$, and $index(v) < i$ otherwise.

To compute the mappings in linear time, data structures $set(i)$,
$size(s)$ and $j$ are maintained during the process of building
$\alpha$. For each $s$, $size(s)$ is the count of assigned vertices
in $s$ if $s$ is nonexhausted and $-1$ otherwise. For $i \in
0..n-1$, $set(i)$ is the set of nonexhausted edges that have exactly
$i$ assigned vertices by $\alpha$. Index $j$ is the maximum $i$ such
that $set(i)$ is nonempty.

\begin{algorithm}[htb!]
\setlength{\algomargin}{3em}

\defKeywords

\dontprintsemicolon
   % \fun{\isAcyclic({\bf in} $(\nodes, \edges)$, {\bf out} $R$, $\gamma$)}{
   \fun{\isAcyclic({\bf in} $\edges$, {\bf out} $R$, $\gamma$)}{
    \nl Let $n$ be the number of nodes in $U(\edges$) \;
    \nl \For{ each $i \in 0..n-1$}{
    \nl        $set(i) \leftarrow \emptyset$ \;
        }
    \nl \For{$E \in \edges$}{
    \nl     $size(E) \leftarrow 0$ \;
    \nl     $\gamma(E) \leftarrow undefined$ \;
    \nl     add $E$ to $set(0)$ \;
        }
    \nl $i \leftarrow n+1, j \leftarrow 0, k \leftarrow 0$ \;
    \nl \While{$j \ge 0$}{
    \nl     delete any $E$ from $set(j)$ \;
    \nl     $k++$ \;
    \nl     $\beta(E) \leftarrow k, R(k) \leftarrow E, size(E)
    \leftarrow -1$ \;
    \nl     \For{ $v \in E$ such that $\alpha(v)$ is not assigned}{
    \nl         $i--$ \;
    \nl         $\alpha(v) \leftarrow i, \beta(v) \leftarrow k$ \;
    \nl         \For{$F \in \edges$ such that $v \in F$ \AND $size(F) \ge 0$}{
    \nl             $\gamma(F) \leftarrow k$ \;
    \nl             delete $F$ from $set(size(F))$ \;
    \nl             $size(F)++$ \;
    \nl             \If{$size(F) < |F|$}{ \;
    \nl                 add $F$ to $set(size(F))$ \;
    \nl                 \If{$j < size(F)$}{
    \nl                     $j \leftarrow size(F)$ \;
                        }
                    } \Else{ \;
    \nl                 $size(F) \leftarrow -1$ \;
                    }
                }
            }
    \nl     \lWhile{$j \ge 0$ \AND $set(j) = \emptyset$}{ $j--$} \;
        }

%    \nl \For{each $i \in 1..k$}{
%    \nl     \For{each $E \in \edges$ such that $\gamma(E) = i$}{
%    \nl         \If{ {\bf not} ($E \cap \{\}$)} {
%    \nl             \Ret \False \;
%                }
%            }
%        }

    \nl \lFor{$v \in U(\edges)$}{$index(v) \leftarrow 0$} \;
    \nl \For{each $i \in 1..k$}{
    \nl     \lFor{$v \in R(i)$}{$index(v) \leftarrow i$} \;
    \nl     \For{each $E \in \edges$ such that $\gamma(E) = i$}{
    \nl         \For{ $v \in E$} {
    \nl             \If{$\beta(v) < i$ \AND $index(v) < i$}{
    \nl                 \Ret \False \;
                    }
                }
            }
        }
        \Ret \True \;
   }
   ------------  \genForest  -----------\;
   \fun{\genForest({\bf in} $\edges$, $R$, $\gamma$, {\bf out} $\mathcal{T}$)}{
     \nl $V \leftarrow \edges$ \;
     \nl $E \leftarrow \{ \{F, R(\gamma(F))\} ~|~ $ \;
            $\hspace{1cm} F \in \edges \mbox{ and } \gamma(F) \mbox{ is
     defined}\}$ \;
     \nl $\mathcal{T} \leftarrow (V, E)$
   }
\caption{\Indp \scriptsize Acyclicity test and generation of the
forest} \label{alg:acyclicity}
\end{algorithm}

The algorithms to test acyclicity and generate the forest are of
linear time complexity \citep{TarjanY84}. Hence, we have the
following result.

\begin{theorem}
The worst case time complexity of the algorithm to identify the tree
convexity of a collection of sets is linear in the problem size.
\end{theorem}

Given a collection of sets $S = \{S_1, S_2, \cdots, S_m\}$, the size
of the problem is $\Sigma_{i=1}^m(|S_i|)$. The complexity of the
acyclicity based algorithm is linear to the problem size. Conitzer
et al.'s algorithm has a complexity of $O(mn^2)$ where $n = |\cup
S|$. Note that the size of each set (bid) may range from $1$ to $n$,
but never exceeds $n$. So, the difference of the worst case
complexity of the two algorithms is clear.

Algorithm~\ref{alg:acyclicity} differs from that of \citep{TarjanY84}
in the following two parts. 1) Line 14 was $i++$ in the original
paper, which was clearly a typo. 2) Instead of having line 22-23,
the original algorithm increases $j$ by one right before line 25,
which is not correct. Our newly added code in line 22-23 will
preserve the linear complexity of the algorithm. In the complexity
analysis, line 25 is the key. The number of executions of line 25
during the whole process can be taken as a combination of two parts:
executions caused by the monotonic decrease of $j$, and those extra
executions $d$ caused by the increase of $j$ in line 22-23. $d$ is
$n$ in the worst case as every node of $U(\edges)$ will be selected
once and only once and for each selected node $d$ will be increased
by only one in the worst case. The new change follows the
amortization spirit used in the original analysis. Therefore,
Algorithm~\ref{alg:acyclicity} still has linear complexity.

In the following comment, we use the notations and refer to the
original algorithm (page 573) in \citep{TarjanY84}. In a personal
communication, Yanakakis and Tarjan points out two alternatives to
correct the original algorithm. The first is to replace $j:=j+1$ by
$j:=|R(k)|$. The other way is to move $j:=j+1$ to the line
immediately before the inner for loop. , i.e., line 15, where $i$ is
updated.
%(In our implementation, all these corrections have similar
% performance.)

\section{Experimental evaluation }
\label{sec:experiments} We have carried out an experimental
evaluation of the performance of the acyclicity based algorithm and
the spanning tree based algorithm \citep{ConitzerDS04}. The algorithm
in \citep{ConitzerDS04} consists of two parts: the first part is to
find a tree over the items (see the background section) and the
second part is to test whether every set (bid) is a subtree of the
constructed tree. Due to space limitation, no concrete algorithm for
the second part is provided in \citep{ConitzerDS04}. However, it is
mentioned in \citep{ConitzerDS04} that the missed algorithm is
achievable in $O(mn)$ where $m$ is the number of sets (bids), and
$n$ the number of elements (items). To make this paper complete and
the experiments here reproducible, we include an algorithm for the
second part. The idea is to get the subgraph of the tree induced
from each set (line 1-4) and then check the connectedness of each
induced graph (line 5-6).

\begin{algorithm}[htb!]
\setlength{\algomargin}{3em}

\defKeywords

\dontprintsemicolon
   \fun{\treeTest({\bf in} $S$, $T$)}{
     \nl \lFor{each $s \in S$}{ construct graph $G_s = (s, \emptyset)$} \;
     \nl \For{each edge $\{a,b\}$ of $T$}{
     \nl    \For{each $s \in S$}{
     \nl        \If{$\{a, b\} \in s$}{
                    add edge $\{a, b\}$ to graph $G_s$
                }
            }
         }
     \nl \For{each graph $G_s$}{
     \nl    \If{the connected component of $G_s$ is not equal to
     $s$}   {
                \Ret \False
            }
         }
     \nl \Ret \True
   }

\caption{\Indp \scriptsize Identify tree convex sets with respect to
a given tree} \label{alg:treeTest}
\end{algorithm}

For line 6, the connected component of a graph can be identified in
linear time \citep{CLR90}. The complexity of the algorithm is $O(mn)$
due to the two loops (line 2 and 3).

Recall that a collection of sets, i.e., a set of bids, is {\em tree
convex} iff there is an {\em item tree} for the bids. So the
algorithm in \citep{ConitzerDS04} is directly applicable to tree
convexity test and thus no modification or reconstruction is
necessary. Our implementation is faithful to the algorithm given in
\citep{ConitzerDS04}. The experiments are carried out on an AMD Opteron 2350 CPU
(frequency 2.0 GHz) with Ubuntu Linux 9.04 of kernel 2.6.28-11. The
algorithms are implemented using Python 2.6.2.

From our implementation, we have the following comments about the
simplicity of the algorithms. Both algorithms are conceptually quite
simple. However, as for implementation, we find that the pseudo code
and data structures of the acyclicity algorithm can be ``directly''
implemented. When we implement the spanning tree based algorithms we
have to choose the data structures on  graphs carefully so that all
the complexity results follow. The final implementation code is much
more complex and longer than that of the acyclicity based algorithm.

Acyclic based and spanning tree based algorithms are evaluated on
random problems (generated by ourselves) and the structured problems
provided by \citet{CATS}.

\subsection{Random problems}
Four parameters are employed to generate our own collections of sets:
$\langle m, n, r_1, r_2 \rangle$ where $m$ denotes the number of
sets of the collection to generate, the size of the sets is between
$r_1$ and $r_2$, and each set takes values from $1$ to $n$.

The evaluation is designed as follows. Since the acyclicity based
algorithm is theoretically faster than the spanning tree based
algorithm, for large problems, its practical performance  should
also be faster. We sample a few problems with large configuration
parameters to show how the difference between these two algorithms
could be. From Table~\ref{largeCase} where the time is for 10
problem instances, the acyclicity based algorithm is one to two
orders of magnitude faster than that of the spanning tree based
algorithm. As the problem size grows, the cost of spanning tree
based algorithm grows much faster than that of the acyclicity based
algorithm.

% m=n=100 r2=10, ours is 0.04s vs existing 1.38s (on
%  10 instances);
% m=n=300, r2=30, 0.31s vs 611.83s; m=n=500 and r2=50, 0.82s vs
% 50min+. We can easily add these results into the future version the
% time elapsed to 500,500,50 is 13742.65s/10 instances
\begin{table}[htb!]
\begin{center}
     \begin{tabular}{| c | c | c | c | c | c |}
%      \hline \multirow{2}{*}{$m$} & \multirow{2}{*}{$n$} & \multirow{2}{*}{$r_1$} & \multirow{2}{*}{$r_2$} & Acyclicity& Spanning tree  \\
%           &  & & & based (s) & based (s) \\
      \hline $m$ & $n$ & $r_1$ & $r_2$ & Acyclicity based & Spanning tree based \\
       \hline  100 & 100 & 2 & 10 & 0.05 & 1.03\\
       \hline  300 & 300 & 2 & 30 & 0.21 & 15.99 \\
       \hline  500 & 500 & 2 & 50 & 0.56 & 69.40 \\
       \hline
     \end{tabular}
\end{center}
\caption{\label{largeCase} Performance for large parameters}
\end{table}

\begin{figure}[htbp]
\begin{center}
\includegraphics[scale=1]{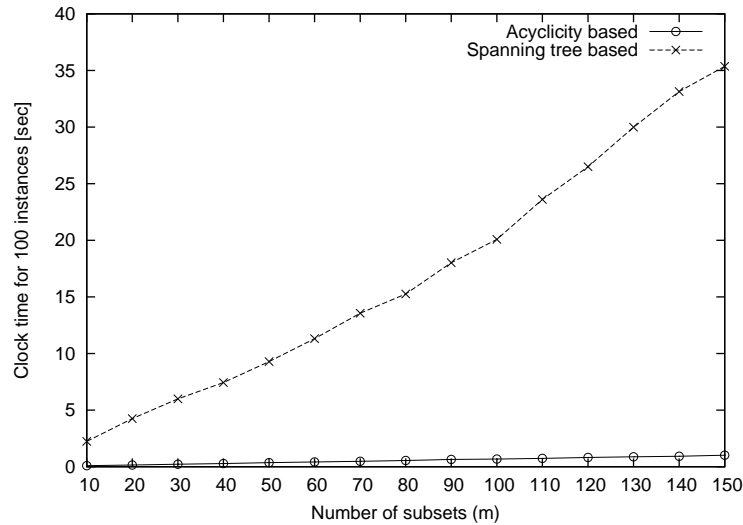} % eps file
\caption{\label{fig:test1}{ Performance of the algorithms on
problems $<m, 100, 2, 30>$ with $m$ changing from 10 to 200 with a
step of 10}}
\end{center}
\end{figure}

\begin{figure}[htbp]
\begin{center}
\includegraphics[scale=1]{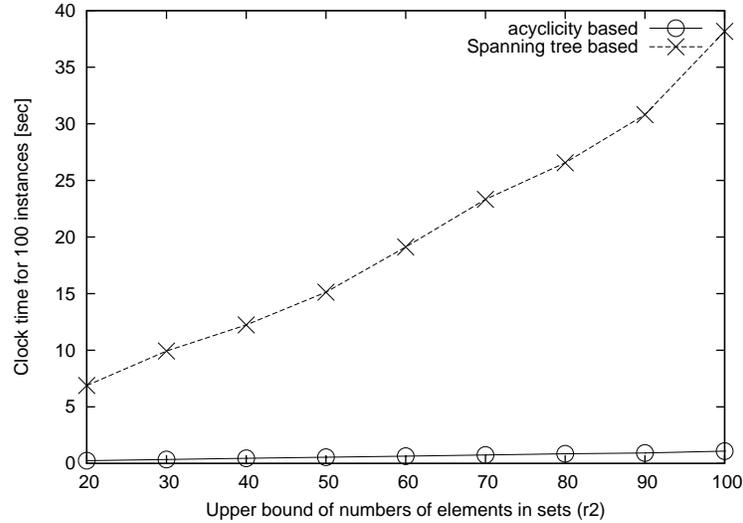}
\caption{\label{fig:test2} { Performance of the algorithms on
problems with $<50, 100, 2, r_2>$ with $r_2$ varying from 20 to 90
with step 10.}}
\end{center}
\end{figure}

For small problems, theoretical time complexity might not fully
agree with practical performance. Therefore, we employ a  systematic
comparison scheme: vary the value of $m$ and $r_2$ respectively with
other parameters fixed. Specifically, we have tested the following
configurations $<m, 100, 2, r_2>$ where $m$ changes from 10 to 200
with a step of 10,  and $r_2$ changes from 20 to 90 with step 10.
100 instances are generated from each configuration of the
parameters. Samples of the results are shown in
Figure~\ref{fig:test1} and Figure~\ref{fig:test2}.

From the results, the acyclicity algorithm runs significantly faster
than the spanning tree based algorithm.

\subsection{Existing structured problems}

The problems \citep{CATS} used in our experiments are {\em arbitrary,
matching, paths, regions, scheduling and Legacy (L1-L8)}. Their
instances are generated from the program at
\url{http://www.cs.ubc.ca/~kevinlb/CATS/}.
% CATS-readme.html
The details of the description of these problems can be found at
\citep{CATS}. Each problem instance is a set of bids. Our task is to
check the tree convexity of the bids. The results are listed in
Table \ref{benchmark}. In the table, each time entry is for 50
instances. From Table \ref{benchmark}, the acyclicity based
algorithm is 30 to 80 times faster than the spanning tree base
algorithm. It is worth of mentioning that all the instances in the
benchmarks are not tree convex, which partially justify our use of
random problems that include both tree convex and non tree convex
instances.

In summary, for both random problems and structured problems, the
acyclicity based algorithm has a clear performance advantage over
the spanning tee based algorithm.

% on processing combinatorial auction problems.

\begin{table}[htb!]
\begin{center}
     \begin{tabular}{| c | c | @{\hskip 3em}r@{.}p{0.2in} |}
%        \hline \multirow{2}{*}{Instance} & Acyclicity  & \multicolumn{2}{c|}{Spanning tree} \\
%          & based (s) & \multicolumn{2}{c|}{based (s)} \\
        \hline Instance & Acyclicity based  & \multicolumn{2}{c|}{\hskip -3em Spanning tree based} \\
       \hline  arbitrary & 0.58 & 34 &  67\\
       \hline  arbitrary-npv & 0.59 & 34 & 01 \\
       \hline  arbitrary-upv & 0.59 & 34 & 91 \\
       \hline  matching & 0.18 & 6 & 14 \\
       \hline  paths & 0.29 & 16 & 27 \\
       \hline  regions & 0.61 & 35 & 19 \\
       \hline  regions-npv & 0.62 & 33 & 68 \\
       \hline  regions-upv & 0.63 & 35 & 37 \\
       \hline  scheduling & 0.17 & 42 & 38 \\
       \hline  L1 & 2.57 & 159 & 84\\
       \hline  L2 & 4.04 & 324 & 02 \\
       \hline  L3 & 0.17 & 8 & 59 \\
       \hline  L4 & 0.16 & 6 & 95 \\
       \hline  L5 & 0.22 & 13 & 76 \\
       \hline  L6 & 0.29 & 18 & 43 \\
       \hline  L7 & 1.61 & 84 & 95 \\
       \hline  L8 & 0.62 & 8 & 8 \\
       \hline
     \end{tabular}
\end{center}
\caption{\label{benchmark} Performance of the algorithms on the
benchmarking problems in \citep{CATS}}
\end{table}

\section{Conclusion}
Polynomial algorithms have been designed to test tree convexity
using ideas from consecutive ones property test and spanning tree.
However, when the collection of sets is taken as a hypergraph, one
can characterize the tree convexity by the acyclicity of the dual
graph of the sets, which leads to a linear test algorithm thanks to
the linear algorithm for testing the acyclicity of hypergraphs. In
addition to its theoretical worst case efficiency, the acyclicity
based algorithm is also very easy to implement and performs very
well compared with the spanning tree based algorithm on the random
problems we have generated.
% This result shows that the identification of tree convex sets is as
%efficient as that of row convex sets.
We notice that the algorithms to test row convexity (i.e.,
consecutive ones property) have been much more involved than the
algorithm to test tree convexity although efforts have been made to
find simpler algorithms \citep{MeidanisPT98,HabibMPV00}. We are not
aware of any work on consecutive ones property employing the
properties of hypergraphs. It is interesting to investigate whether
hypergraph properties and algorithms can help produce efficient and
simple consecutive ones property test algorithms.

\section*{Acknowledgment}
We thank anonymous referees of earlier drafts of this work for
pointing out to us Goodman and Shmueli's results \citep{GoodmanS83},
and Gottlob and Greco's result \citep{GottlobG07}.

\bibliographystyle{abbrvnat}
\bibliography{cp}

\begin{thebibliography}{29}
\providecommand{\natexlab}[1]{#1}
\providecommand{\url}[1]{\texttt{#1}}
\expandafter\ifx\csname urlstyle\endcsname\relax
  \providecommand{\doi}[1]{doi: #1}\else
  \providecommand{\doi}{doi: \begingroup \urlstyle{rm}\Url}\fi

\bibitem[Beeri et~al.(1983)Beeri, Fagin, Maier, and Yannakakis]{BeeriFMY83}
C.~Beeri, R.~Fagin, D.~Maier, and M.~Yannakakis.
\newblock On the desirability of acyclic database schemes.
\newblock \emph{J. ACM}, 30\penalty0 (3):\penalty0 479--513, 1983.
\newblock ISSN 0004-5411.
\newblock \doi{http://doi.acm.org/10.1145/2402.322389}.

\bibitem[Berge(1973)]{Berge73}
C.~Berge.
\newblock \emph{Graphs and Hypergraphs}.
\newblock American Elsevier Publishing Company, 1973.

\bibitem[Bernstein and Goodman(1981)]{BernsteinG81}
P.~A. Bernstein and N.~Goodman.
\newblock Power of natural semijoins.
\newblock \emph{SIAM J. Comput.}, 10\penalty0 (4):\penalty0 751--771, 1981.

\bibitem[Booth and Lueker(1976)]{BoothL76}
K.~S. Booth and G.~S. Lueker.
\newblock Testing for the consecutive ones property, interval graphs, and graph
  planarity using pq-tree algorithms.
\newblock \emph{J. Comput. Syst. Sci.}, 13\penalty0 (3):\penalty0 335--379,
  1976.

\bibitem[Conitzer et~al.(2004)Conitzer, Derryberry, and Sandholm]{ConitzerDS04}
V.~Conitzer, J.~Derryberry, and T.~Sandholm.
\newblock Combinatorial auctions with structured item graphs.
\newblock In \emph{AAAI}, pages 212--218, 2004.

\bibitem[Cormen et~al.(1990)Cormen, Leiserson, and Rivest]{CLR90}
T.~H. Cormen, C.~E. Leiserson, and R.~L. Rivest.
\newblock \emph{Introduction to Algorithms}.
\newblock MIT Press, Cambridge, MA, 1990.

\bibitem[Cramton et~al.(2006)Cramton, Shoham, and Steinberg]{CramtonSS06}
P.~Cramton, Y.~Shoham, and R.~Steinberg.
\newblock \emph{Combinatorial Auctions}.
\newblock MIT Press, 2006.

\bibitem[Dechter(2003)]{Dec03}
R.~Dechter.
\newblock \emph{Constraint Processing}.
\newblock Morgan Kaufmann, San Francisco, CA, 2003.

\bibitem[Dechter and Pearl(1989)]{DP89}
R.~Dechter and J.~Pearl.
\newblock Tree clustering for constraint networks.
\newblock \emph{Artificial Intelligence}, 38:\penalty0 353--366, 1989.

\bibitem[Fagin(1983)]{Fagin83}
R.~Fagin.
\newblock Degrees of acyclicity for hypergraphs and relational database
  schemes.
\newblock \emph{J. ACM}, 30\penalty0 (3):\penalty0 514--550, 1983.

\bibitem[Freuder(1978)]{Fre78}
E.~Freuder.
\newblock Synthesizing constraint expressions.
\newblock \emph{Communications of ACM}, 21\penalty0 (11):\penalty0 958--966,
  1978.

\bibitem[Fulkerson and Gross(1965)]{FulkersonG65}
D.~Fulkerson and O.~Gross.
\newblock Incidence matrices and interval graphs.
\newblock \emph{Pac. J. Math.}, 15:\penalty0 835--855, 1965.

\bibitem[Goodman and Shmueli(1983)]{GoodmanS83}
N.~Goodman and O.~Shmueli.
\newblock Syntactic characterization of tree database schemas.
\newblock \emph{J. ACM}, 30\penalty0 (4):\penalty0 767--786, 1983.

\bibitem[Gottlob and Greco(2007)]{GottlobG07}
G.~Gottlob and G.~Greco.
\newblock On the complexity of combinatorial auctions: structured item graphs
  and hypertree decomposition.
\newblock In \emph{ACM Conference on Electronic Commerce}, pages 152--161,
  2007.

\bibitem[Gottlob and Szeider(2008)]{GottlobS08}
G.~Gottlob and S.~Szeider.
\newblock Fixed-parameter algorithms for artificial intelligence, constraint
  satisfaction and database problems.
\newblock \emph{Comput. J.}, 51\penalty0 (3):\penalty0 303--325, 2008.

\bibitem[Habib et~al.(2000)Habib, McConnell, Paul, and Viennot]{HabibMPV00}
M.~Habib, R.~M. McConnell, C.~Paul, and L.~Viennot.
\newblock {Lex-BFS} and partition refinement, with applications to transitive
  orientation, interval graph recognition and consecutive ones testing.
\newblock \emph{Theor. Comput. Sci.}, 234\penalty0 (1-2):\penalty0 59--84,
  2000.

\bibitem[Hsu(2002)]{Hsu02}
W.-L. Hsu.
\newblock A simple test for the consecutive ones property.
\newblock \emph{J. Algorithms}, 43\penalty0 (1):\penalty0 1--16, 2002.
\newblock ISSN 0196-6774.
\newblock \doi{http://dx.doi.org/10.1006/jagm.2001.1205}.

\bibitem[Leyton-Brown et~al.(2000)Leyton-Brown, Pearson, and Shoham]{CATS}
K.~Leyton-Brown, M.~Pearson, and Y.~Shoham.
\newblock Towards a universal test suite for combinatorial auction algorithms.
\newblock In \emph{Proceedings of 2nd ACM Conference on Electronic Commerce},
  2000.

\bibitem[Mackworth(1977)]{Mac77}
A.~K. Mackworth.
\newblock Consistency in networks of relations.
\newblock \emph{Artificial Intelligence}, 8\penalty0 (1):\penalty0 118--126,
  1977.

\bibitem[Meidanis et~al.(1998)Meidanis, Porto, and Telles]{MeidanisPT98}
J.~Meidanis, O.~Porto, and G.~P. Telles.
\newblock On the consecutive ones property.
\newblock \emph{Discrete Applied Mathematics}, 88\penalty0 (1-3):\penalty0
  325--354, 1998.

\bibitem[Montanari(1974)]{Mon74}
U.~Montanari.
\newblock Networks of constraints: fundamental properties and applications.
\newblock \emph{Information Science}, 7\penalty0 (2):\penalty0 95--132, 1974.

\bibitem[Rose et~al.(1976)Rose, Tarjan, and Lueker]{RoseTL76}
D.~J. Rose, R.~E. Tarjan, and G.~S. Lueker.
\newblock Algorithmic aspects of vertex elimination on graphs.
\newblock \emph{SIAM J. Comput.}, 5\penalty0 (2):\penalty0 266--283, 1976.

\bibitem[Rothkopf et~al.(1998)Rothkopf, Pekec, and Harstad]{RothkopfPH98}
M.~H. Rothkopf, A.~Pekec, and R.~M. Harstad.
\newblock Computationally manageable combinatorial auctions.
\newblock \emph{Management Science}, 44\penalty0 (8):\penalty0 1131--1147,
  1998.

\bibitem[Sandholm and Suri(2003)]{SandholmS03}
T.~Sandholm and S.~Suri.
\newblock Bob: Improved winner determination in combinatorial auctions and
  generalizations.
\newblock \emph{Artif. Intell.}, 145\penalty0 (1-2):\penalty0 33--58, 2003.

\bibitem[Tarjan and Yannakakis(1984)]{TarjanY84}
R.~E. Tarjan and M.~Yannakakis.
\newblock Simple linear-time algorithms to test chordality of graphs, test
  acyclicity of hypergraphs, and selectively reduce acyclic hypergraphs.
\newblock \emph{SIAM J. Comput.}, 13\penalty0 (3):\penalty0 566--579, 1984.

\bibitem[van Beek and Dechter(1995)]{vBD95}
P.~van Beek and R.~Dechter.
\newblock On the minimality and global consistency of row-convex constraint
  networks.
\newblock \emph{Journal of The ACM}, 42\penalty0 (3):\penalty0 543--561, 1995.

\bibitem[Wallace(1993)]{Wal93}
R.~J. Wallace.
\newblock Why {AC-3} is almost always better than {AC-4} for establishing arc
  consistency in {CSPs}.
\newblock In \emph{Proceedings of IJCAI-93}, pages 239--247, Chambery, France,
  1993. IJCAI Inc.

\bibitem[Yosiphon(2003)]{Yos03}
G.~Yosiphon.
\newblock Efficient algorithm for identifying tree convex constraints.
\newblock Manuscript, 2003.

\bibitem[Zhang and Yap(2003)]{ZY03b}
Y.~Zhang and R.~H.~C. Yap.
\newblock Consistency and set intersection.
\newblock In \emph{Proceedings of International Joint Conference on Artificial
  Intelligence 2003}, pages 263--268, Acapulco, Mexico, 2003. IJCAI Inc.

\end{thebibliography}

\end{document}